\documentclass[%
letter,
reprint,
 amsmath,amssymb,
 aps,
]{revtex4-1}

\usepackage{graphicx}
\usepackage{dcolumn}
\usepackage{bm}

\begin{document}

\preprint{APS/123-QED}

\title{A commend on ``Three Classes of Newtonian Three-Body Planar Periodic Orbits''  \\  by   \v{S}uvakov and Dmitra\v{s}inovi\'{c} (PRL, 2013) }

\author{Xiaoming Li}

\affiliation{School of Naval Architecture, Ocean and Civil Engineering\\
Shanghai Jiaotong University, Shanghai 200240, China
}

\author{Shijun Liao}
\email{sjliao@sjtu.edu.cn}

\affiliation{School of Naval Architecture, Ocean and Civil Engineering\\
Key Laboratory of Education-Ministry in Scientific Computing\\
Shanghai Jiaotong University,  Shanghai 200240,  China\\
}

\date{\today}

\begin{abstract}
Currently, the fifteen new periodic solutions of Newtonian three-body problem  with equal mass were reported by  \v{S}uvakov and Dmitra\v{s}inovi\'{c} (PRL, 2013) \cite{3body-2012}. However,  using a reliable numerical approach (namely the Clean Numerical Simulation, CNS) that is based on the  arbitrary-order Taylor series method and data in arbitrary-digit precision,  it is found that at least seven of them greatly depart from the periodic orbits after a long enough interval of time.  Therefore, the reported initial conditions of at least seven of the  fifteen orbits  reported  by  \v{S}uvakov and Dmitra\v{s}inovi\'{c}  \cite{3body-2012} are  not accurate enough to  predict periodic orbits.  Besides, it is found that  these seven orbits are unstable.
\end{abstract}

\maketitle



According to  H.  Poincar\'{e},  orbits of the famous three-body problem \cite{Valtonen2005} are not integrable in general cases.  Although chaotic orbits of three-body problems widely exist,  three families of  periodic orbits were found:

\begin{enumerate}
\item the Lagrange-Euler family,   dating back to  the analytical solutions in the 18th century (one recent orbit was given by Moore \cite{Moore1993});

\item the Broucke-Hadjidemetriou-H\'{e}non family,  dating back to the mid-1970s \cite {Broucke1975,  Christides1975,  Hadjidemetriou1975,  Hadjidemetriou1975B,  Henon1976, Henon1977};

\item the Figure-8 family,  discovered  in 1993 by Moore \cite{Moore1993}  and  extended to the rotating cases \cite{Nauenberg2001, Chenciner2005, Broucke2006, Nauenberg2007}.

\end{enumerate}
Note that nearly all of these reported periodic orbits are planar.   In 2013, \v{S}uvakov  and Dmitra\v{s}inovi\'{c}  \cite{3body-2012}  found by means of numerical approach that there exist four classes of  planar  periodic orbits of Newtonian three body with equal mass, with  the above three families belonging to  one class.   Besides,   they  reported  three new classes of   planar periodic orbits  and  gave a few initial conditions for each class in 5-digit precision:   they refined
their initial conditions to the level of return proximity of less then $10^{-6}$ by using the gradient descent method.   For the details of  their  15  planar periodic orbits, please refer to the gallery  \cite{gallery}.  Especially,  \v{S}uvakov  and Dmitra\v{s}inovi\'{c}  \cite{3body-2012}   expected their solutions ``to be either stable or marginally unstable, as otherwise they probably would not have been found''  by their numerical method.    

Let the vector ${\bf r}_i(t)$ denote the orbit of three body with equal mass, where $i=1,2,3$,  and $t$ denotes the time, respectively. If the orbit is periodic with the period $T$, it holds $ {\bf r}_i(t)={\bf r}_i(t+nT)$ for {\em arbitrary} time $t\geq 0$ and {\em arbitrary} integer $n\geq 1$.  If,  given a tiny disturbance (for example at $t=0$),  the three-bodies greatly depart their periodic orbits after a long  enough time, then the corresponding periodic orbits are  unstable.

\v{S}uvakov and Dmitra\v{s}inovi\'{c}  \cite{3body-2012} used the gradient descent method to search for the initial conditions of the periodic orbits of three-body with equal mass.   It is well-known that  orbits of three-body problem are often chaotic, i.e. very sensitive to initial conditions.
Thus, it is very important to gain reliable orbits of the three-body problem.  However,  \v{S}uvakov and Dmitra\v{s}inovi\'{c}  \cite{3body-2012} employed it only in the normal precision (i.e. 5-digit).  Thus, their reported periodic orbits should be checked carefully using a more reliable approach.   

To gain mathematically reliable numerical simulations of orbits of Newtonian three body problem, we use the so-called ``Clean Numerical Simulation'' (CNS) \cite{Liao2013, Liao2009}  that is based on arbitrary-order Taylor series method (TSM) \cite{Corliss1982, Barrio2005} and the arbitrary precision library \cite{Oyanarte1990} of C.  The TSM can trace back to Newton, Euler, Liouville and Cauchy.  It has an advantage that its formula at arbitrarily high order can be easily expressed in the same form. So, from viewpoint of numerical simulations, it is rather easy to use the TSM at very high order so as to deduce the truncation error to a required level.  Besides, the round-off  error can be reduced to arbitrary level by means of the multiple precision (MP) library \cite{Oyanarte1990}. Let $M$ denote the order of TSM and $N_s$ the number of significant digits of multiple-precision data, respectively.  Unlike other numerical approaches,  the CNS enforces that $M$ increases together with $N_s$, such as $M =2 N_s$ as illustrated by Liao \cite{Liao2013} for chaotic solutions of Lorenz equation.  More importantly, the reliability of one CNS simulation in a given finite but long enough interval is guaranteed by means of other better CNS simulations using larger $M$ and/or smaller time step $\Delta t$. In this way, the numerical noises can be decreased to such a small level that both truncation and round-off errors are negligible in a given finite but long enough interval.   For example, Liao \cite{Liao2013} employed the CNS to {\em accurately} and {\em reliably} simulate the propagation of  physical uncertainty of initial positions (at the dimensionless level $10^{-60}$) of the chaotic Hamiltonian H\'{e}non-Heiles system for motions of stars in a plane about the galactic center.  Besides,  using 1200 CPUs of the National Supercomputer TH-A1 and the modified parallel integral algorithm based on the CNS with the 3500th-order TSM and the 4180-digit multiple-precision data, Liao and Wang \cite{LiaoWang2013} currently gain, for the first time,  a mathematically reliable simulation of chaotic solution of Lorenz equation in a rather long interval [0,10000].  All of these indicate that the CNS can indeed provide us a safe way to gain mathematically reliable simulations of chaotic dynamic systems in a finite but long enough interval.  All numerical simulations reported here are obtained by the CNS with high enough order of TSM and accurate enough multiple-precision data, whose validity in a given long enough interval is confirmed by other better CNS simulations with higher-order TSM, and/or more accurate MP data,  and/or smaller time step $\Delta t$.   For the detailed numerical algorithm, please refer to Liao \cite{Liao2013B}.


 \v{S}uvakov and Dmitra\v{s}inovi\'{c} \cite{3body-2012} reported the initial conditions of the newly found periodic orbits in the 5-digit precision.  Currently, they obtained the more accurate initial conditions in the 15-digit precision (the seven among them are listed in Table~\ref{table:initial}) for the periodic orbits.   In general, for chaotic dynamic systems, exact initial conditions of the periodic orbits should be irrational numbers, as illustrated by  Viswanath \cite{Viswanath2004} who reported the initial conditions of  periodic solutions of Lorenz equation in accuracy of 500 significant digits.  

\begin{table*}
\caption{The initial conditions of the 7 periodic orbits in 15-digit precision \footnote{These values are given by  M. \v{S}uvakov  in an email communication}. \label{table:initial}}
\begin{ruledtabular}
\begin{tabular}{lccc}
\textrm{Class, number, name} &
$\dot{x}_1(0)$ &
$\dot{y}_1(0)$ & $T$\\
\colrule
I.A.1 butterfly I &  0.306892758965492  & 0.125506782829762 & 6.23564136316479 \\
I.B.4 moth III  &   0.383443534851074   &  0.377363693237305 & 25.8406180475758 \\
I.B.5 goggles   &   0.0833000564575194 &  0.127889282226563 &  10.4668176954385  \\
I.B.7 dragonfly &   0.080584285736084  & 0.588836087036132 &  21.2709751966648  \\
II.B.1 yarn &   0.559064247131347 &  0.349191558837891 &  55.5017624421301  \\
II.C.2a  yin-yang I & 0.513938054919243  & 0.304736003875733&  17.328369755004 \\
II.C.2b yin-yang I &  0.282698682308198 &  0.327208786129952 & 10.9625630756217 \\
\end{tabular}
\end{ruledtabular}
\end{table*}

\begin{table}
\caption{The position $(x_1,y_1)$ of Body-1 at $t=200$ in case of BUTTERFLY-I given by the different orders of TSM  and 300-digit multiple-precision data with the different time steps.  The initial condition is listed in Table~\ref{table:initial}}. \label{table:xy-body1}
\begin{ruledtabular}
\begin{tabular}{ccll}
\textrm{Order} &   $\Delta t$    &
${x}_1(200)$ &
${y}_1(200)$ \\
\colrule
20  & $10^{-5}$     & -33.498137     &   -12.017376  \\
25  & $10^{-5}$     &  -33.498137957  &   -12.017376712  \\
30  & $10^{-5}$     &  -33.49813795772  &   -12.01737671265  \\
40  &  $10^{-5}$     & -33.49813795771996  &   -12.01737671265596  \\
45  &  $10^{-5}$     & -33.49813795771996  &   -12.01737671265596  \\
50  &  $10^{-5}$     & -33.49813795771996  &   -12.01737671265596  \\
12  &  $10^{-6}$     & -33.49813795771996  &   -12.01737671265596  \\
15  &  $10^{-6}$     & -33.49813795771996  &   -12.01737671265596  \\
20  &  $10^{-6}$     & -33.49813795771996  &   -12.01737671265596  \\
12  &  $10^{-7}$     & -33.49813795771996  &   -12.01737671265596  \\
\end{tabular}
\end{ruledtabular}
\end{table}

\begin{figure}[h]
\includegraphics[scale=0.3]{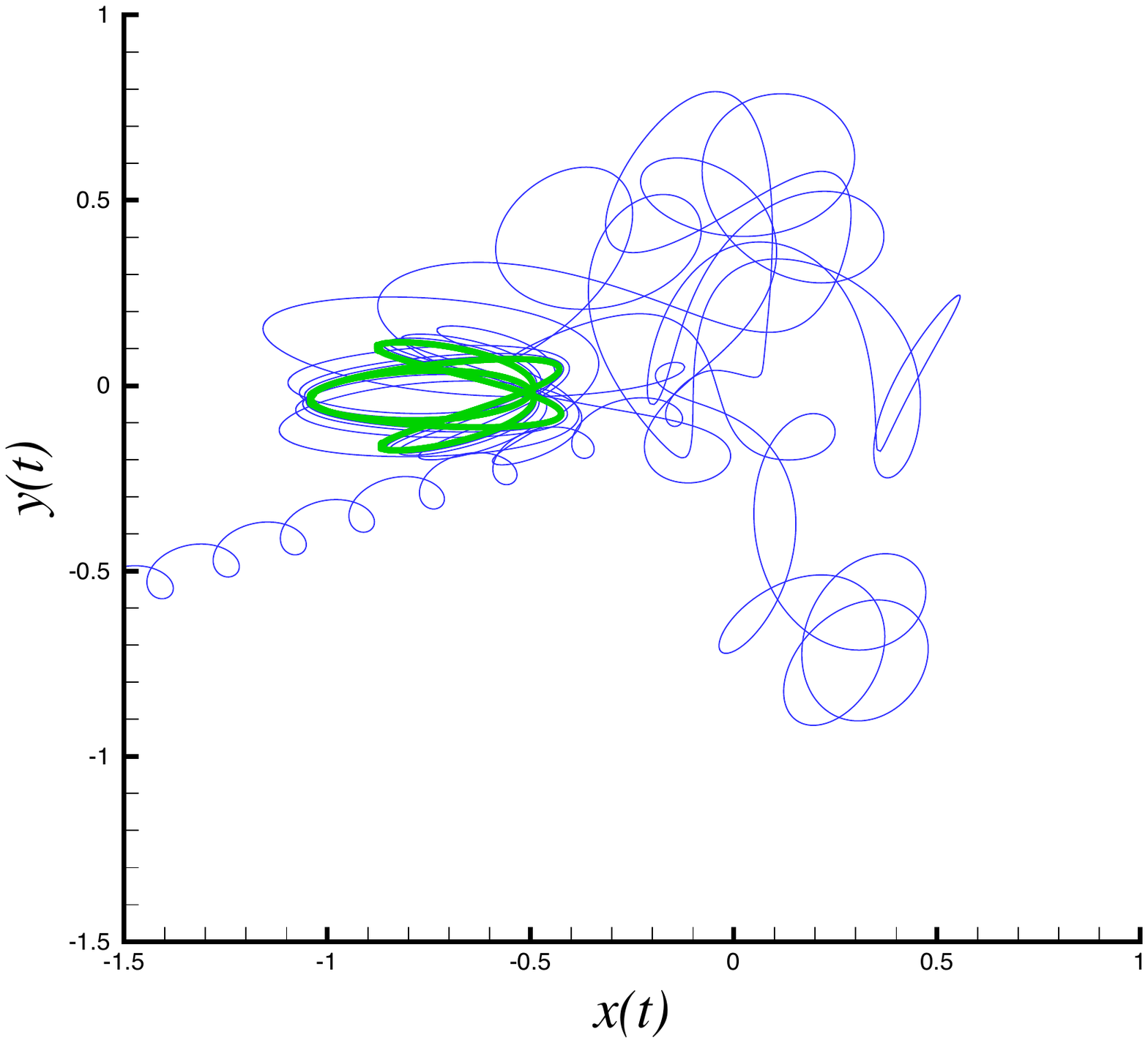}
\caption{Orbits of Body-1 in case of BUTTERFLY-I in the interval [0,200] gained by means of the CNS using 40th-order TSM and 300-digit multiple precision with time step $\Delta t=10^{-5}$. The initial condition in 15-digit precision in Table~\ref{table:initial} is used. Greed line: periodic orbit reported in \cite{3body-2012}.}
\label{Fig:butterfly-I-body1}
\end{figure}

\begin{figure}[h]
\includegraphics[scale=0.3]{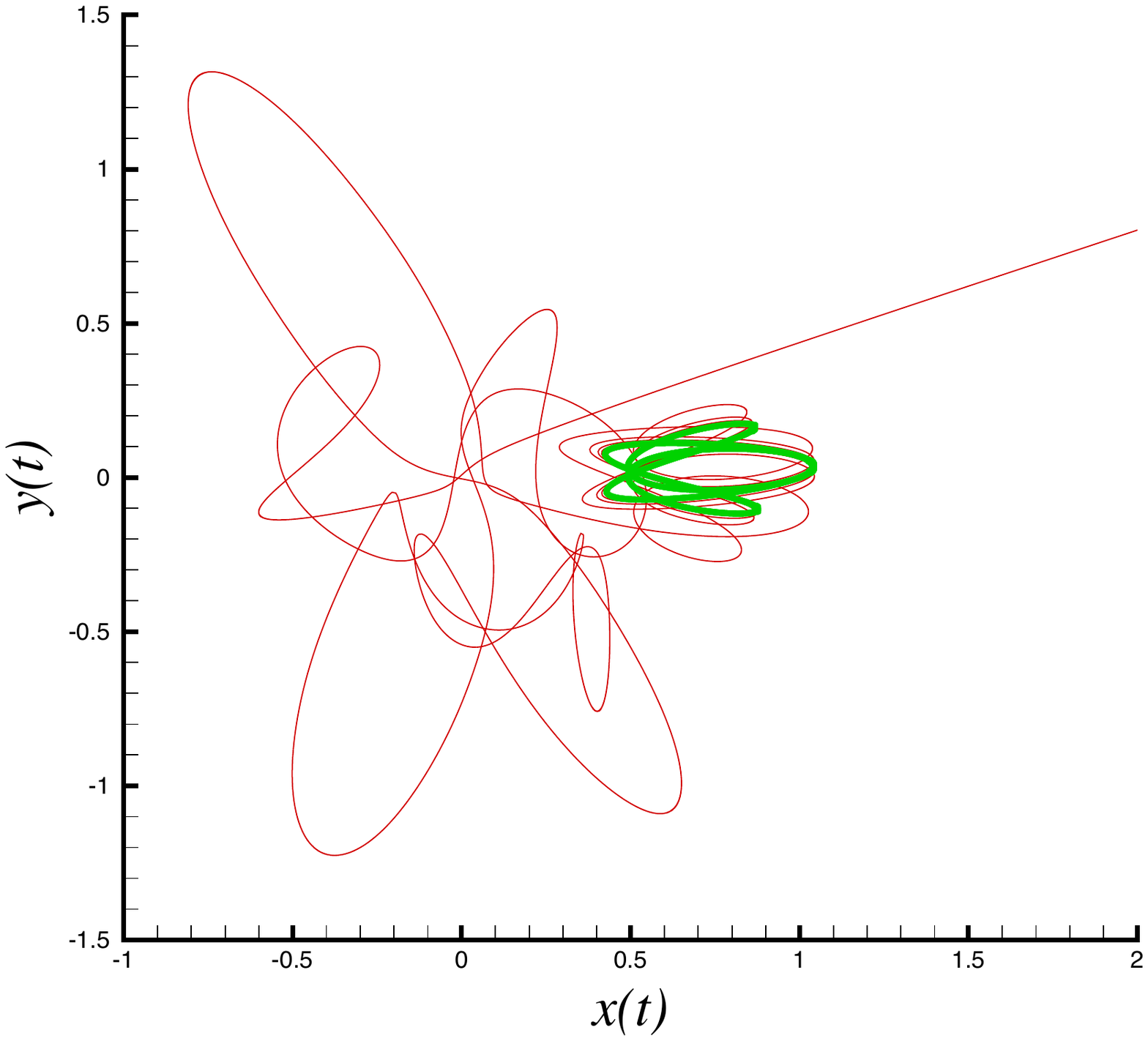}
\caption{Orbits of Body-2 in case of BUTTERFLY-I in the interval [0,200]  gained by means of the CNS using 40th-order TSM condition in 15-digit precision in Table~\ref{table:initial} is used. Greed line: periodic orbit reported in \cite{3body-2012}. }
\label{Fig:butterfly-I-body2}
\end{figure}

\begin{figure}[h]
\includegraphics[scale=0.3]{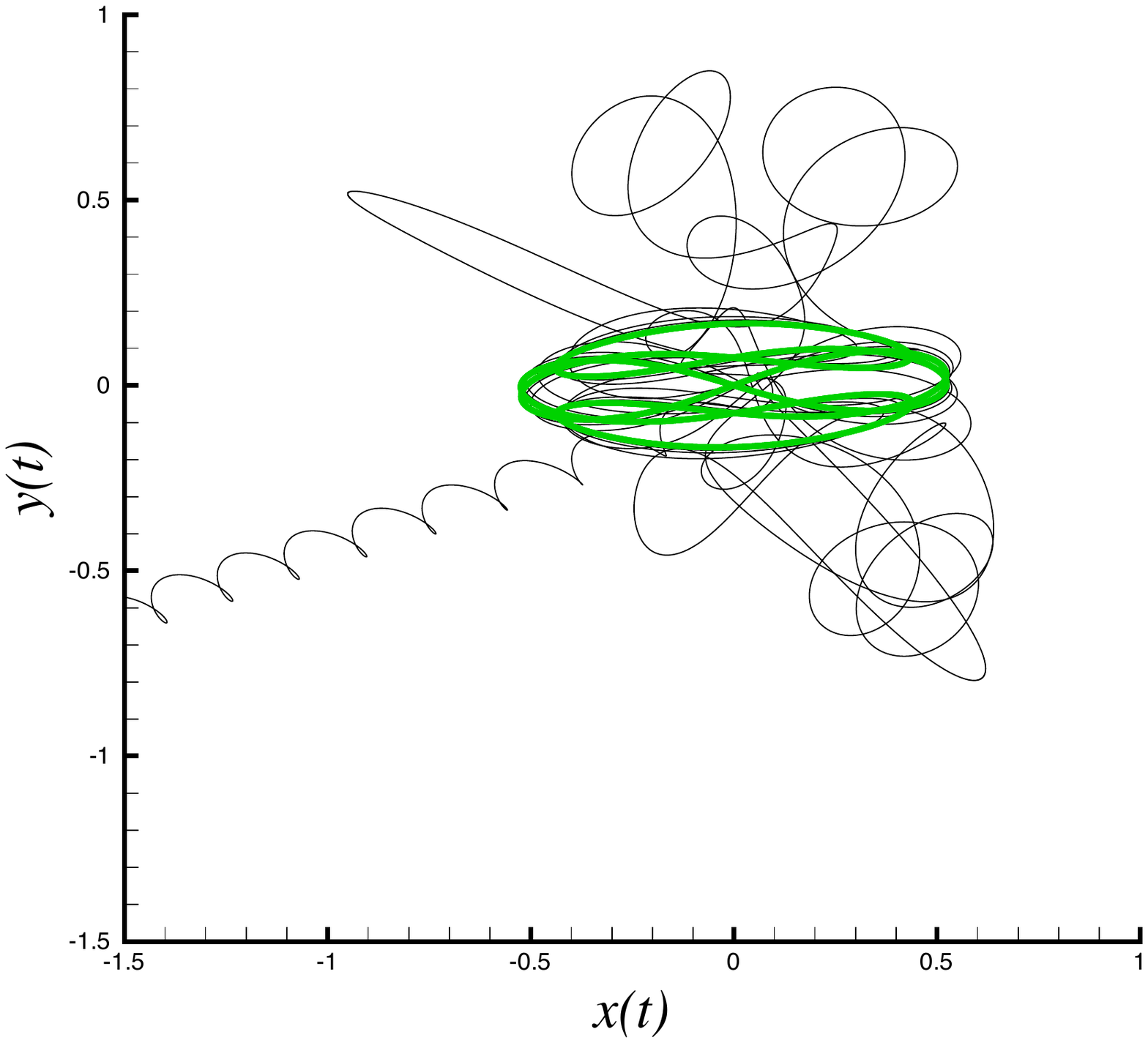}
\caption{Orbits of Body-3 in case of BUTTERFLY-I in the interval [0,200]  gained by means of the CNS using 40th-order TSM and 300-digit multiple precision with time step $\Delta t=10^{-5}$. The initial condition in 15-digit precision in Table~\ref{table:initial} is used. Greed line: periodic orbit reported in \cite{3body-2012}. }
\label{Fig:butterfly-I-body3}
\end{figure}

\begin{figure}
\includegraphics[scale=0.3]{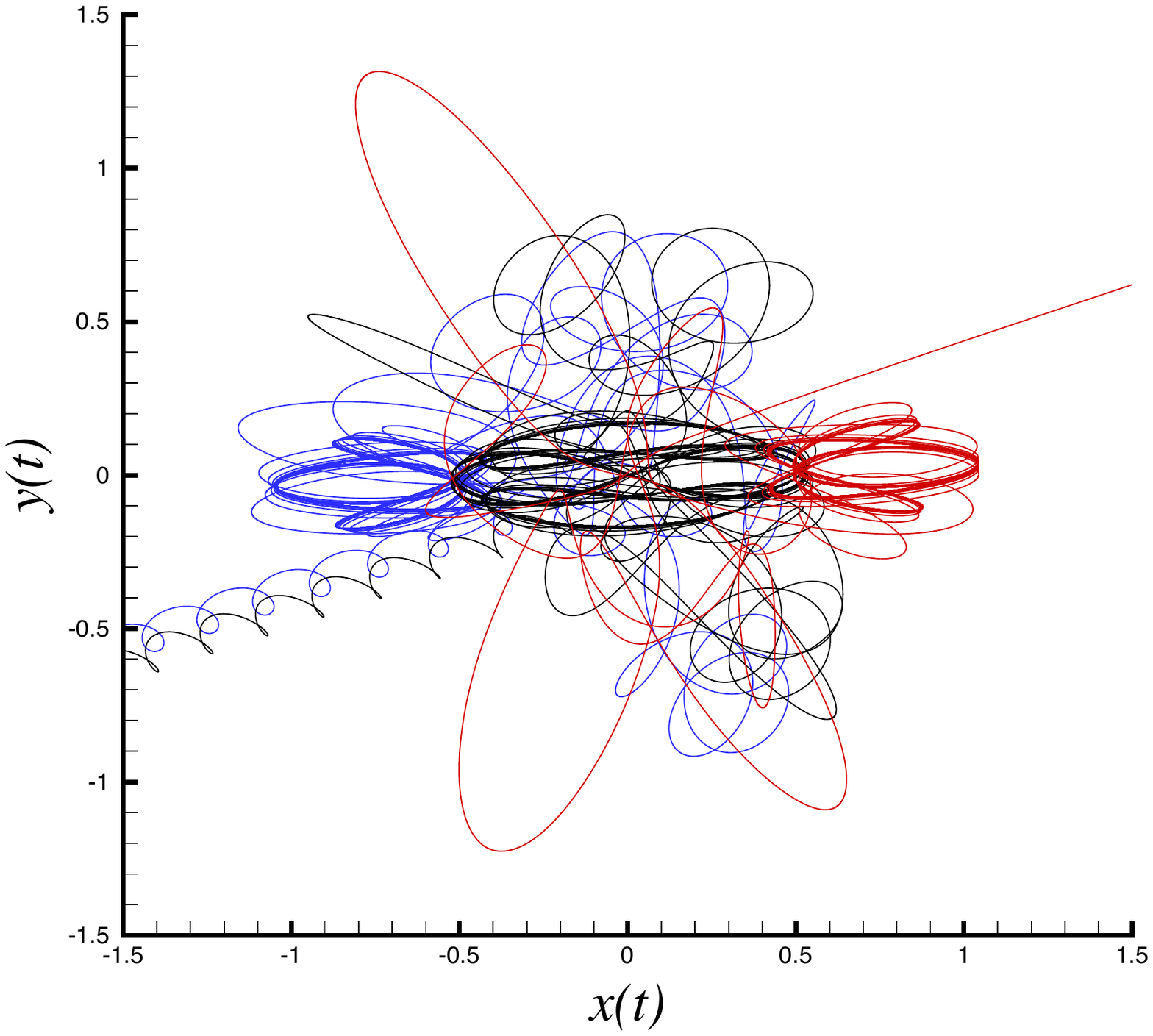}
\caption{Orbits of three bodies in case of BUTTERFLY-I in the interval [0,200]  gained by means of the CNS using 40th-order TSM and 300-digit multiple precision with time step $\Delta t=10^{-5}$. The initial condition in 15-digit precision in Table~\ref{table:initial} is used. Blue line: orbit of Body-1; Red line: orbit of Body-2; Black line: orbit of Body-3.}
\label{Fig:butterfly-I}
\end{figure}

\begin{figure}
\includegraphics[scale=0.3]{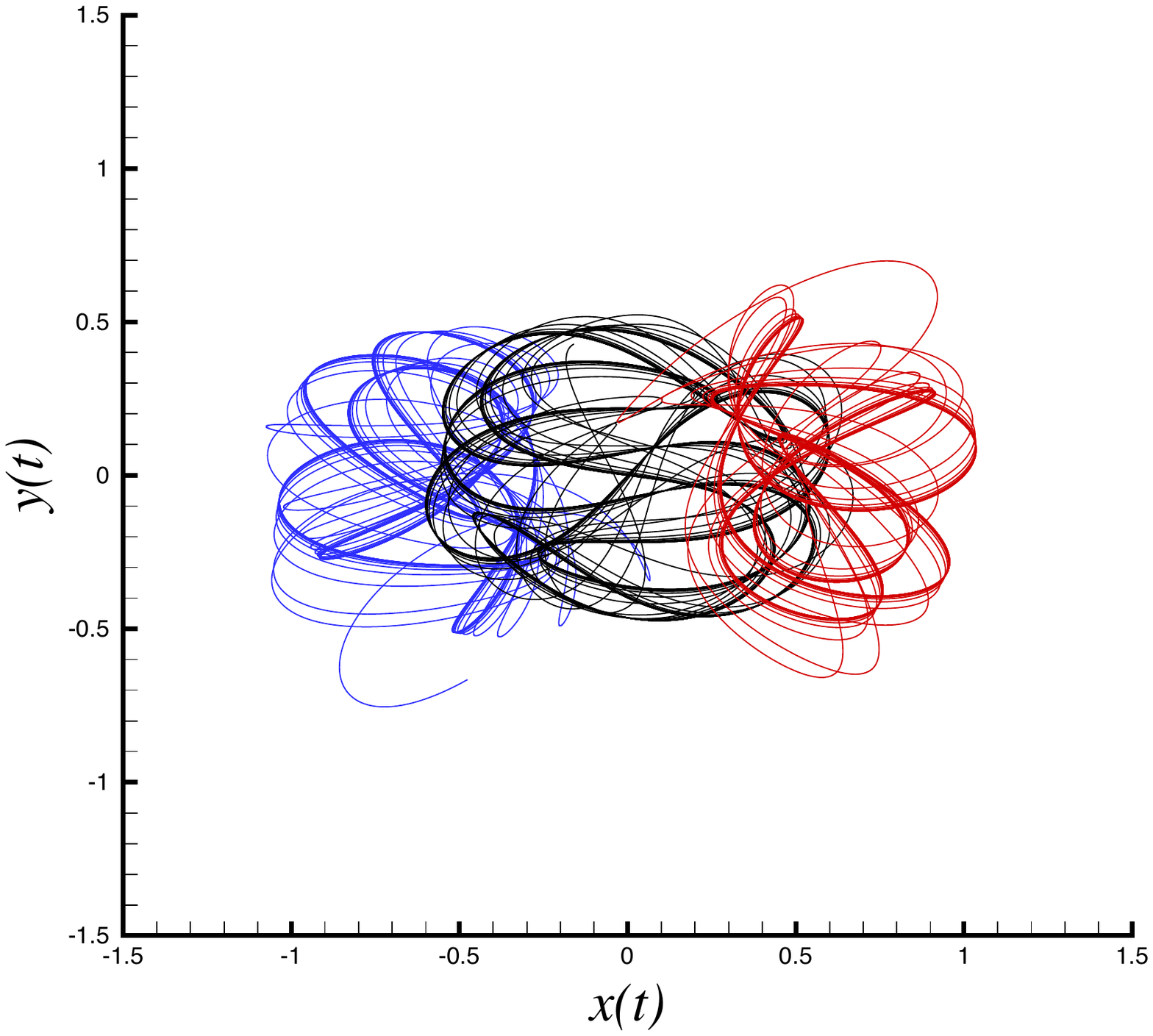}
\caption{Orbits of three bodies in case of II.C.2b (yin-yang I) in the interval [0,190]  gained by means of the CNS using 25th-order TSM and 300-digit multiple precision with time step $\Delta t=10^{-5}$. The initial condition in 15-digit precision in Table~\ref{table:initial} is used. Blue line: orbit of Body-1; Red line: orbit of Body-2; Black line: orbit of Body-3.}
\label{Fig:yinyang1b}
\end{figure}

Without loss of generality, let us consider the case of BUTTERFLY-I, i.e. the Class I.A.1 defined in \cite{3body-2012}.  Using the 20th to 50th order TSM and the 300-digit multiple precision data with the time step $\Delta t = 10^{-5}$, we obtain the corresponding trajectories of the three bodies in the interval [0,200]. It is found that all of these trajectories agree well in the whole interval [0,200]  at least in 7 digits, for example as shown in Table~\ref{table:xy-body1} for the position of Body-1 at $t=200$.  Note that, for arbitrary $M\geq 40$,  the CNS results given by the $M$th-order TSM with $\Delta t=10^{-5}$ have at least the 14 significant digits in the whole interval [0,200], whose reliability is confirmed by using  the $M'$th-order ($M'\geq 12$) TSM with a smaller time step  $\Delta t=10^{-6}$. Therefore, all of these CNS numerical simulations are {\em convergent} to the same result in the interval [0,200] and thus are reliable mathematically.  However, the orbits of the three bodies are almost periodic only up to about $t = 130$, but thereafter depart the periodic ones far and far away, as shown in Figures~\ref{Fig:butterfly-I-body1} to \ref{Fig:butterfly-I-body3}.  According to our reliable simulations in the interval [0,200], we are quite sure that the orbits are completely non-periodic after $t>130$, as shown in Figure~\ref{Fig:butterfly-I}: Body-1 and Body-3 escape together  to become a binary-body system, while Body-2 escapes in the opposite direction.  This counterexample clearly indicates that the given initial condition in 15-digit precision of the periodic orbit BUTTERFLY-I found by \v{S}uvakov and Dmitra\v{s}inovi\'{c} \cite{3body-2012} is {\em not} accurate enough to guarantee periodic orbits.

The reliable simulations of orbits gained by means of the seven initial conditions listed in Table~\ref{table:initial} and the CNS with the 300-digit multiple precision data (we will not repeat this point thereafter), high enough orders of TSM and small enough time step  are gained in a similar way, respectively.  All of these numerical simulations are guaranteed to be reliable in a finite but long enough interval.  However, it is found that all of the seven initial conditions can not guarantee a periodic orbit.

In case of the moth-III, the orbits in the interval [0,590] gained by means of the initial condition listed in Table~\ref{table:initial} and the 20th-order TSM with the 300-digit multiple precision and the time step $\Delta t =10^{-5}$ agree well in the 13 significant digits with those gained by the 25th-order TSM and the same time step.  The orbits are almost periodic up to $t = 560$, i.e. about 22  periods ($T = 25.8406180475758$) of the moth-III, but thereafter depart the periodic ones far and far away.  According to the CNS results given by the 25th-order TSM and the same time step, the collision occurs at about $t=699.45068$, a little more than 27 periods of the periodic orbit reported by \v{S}uvakov and Dmitra\v{s}inovi\'{c} \cite{3body-2012}.  

In the case of the goggles, the orbits in the interval [0,90] given by the CNS with the  30th-order TSM and the time step $\Delta t= 10^{-5}$ agree well in the 8 significant digits with those by the 15th-order TSM and the smaller time step $\Delta t=10^{-6}$.  It is found that the orbits are almost periodic only up to $t = 55$ (i.e. a little more than 5 periods of the goggles reported in \cite{3body-2012}), but thereafter depart from the periodic ones far and far away.

In the case of dragonfly, the orbits in the interval [0,950] given by the CNS with the 25th-order TSM and the time step $\Delta t=10^{-5}$ agree well in the 4 significant digits with those given by the 30th-order TSM and the same time step.  The orbits are almost periodic only up to $t=720$, and thereafter depart the periodic ones far and far away.

In the case of yarn, the orbits in the interval [0,560] given by the CNS with the 25th-order TSM and the time step $\Delta t=10^{-5}$ agree well in the 7 significant digits with those given by the 30th-order TSM and the same time step.   The orbits are almost periodic only up to $t=440$, and thereafter depart the periodic ones far and far away.

In the case of yin-yang I (II.C.2a), the orbits in the interval [0,320] given by the CNS with the 25th-order TSM and the time step $\Delta t=10^{-5}$ agree well in the 16 significant digits with those given by the 30th-order TSM and the same time step.  The orbits are almost periodic only up to $t=250$, thereafter depart from the periodic ones far and far away.

In the case of yin-yang I (II.C.2b), the orbits in the interval [0,190] given by the CNS with the 25th-order TSM and the time step $\Delta t=10^{-5}$ agree well in the 6 significant digits with those given by the 30th-order TSM and the same time step.  The orbits are almost periodic only up to $t=135$, thereafter depart from the periodic ones far and far away, as shown in Fig.~\ref{Fig:yinyang1b}. 

Besides, it is found that the orbits are sensitive to the initial conditions: adding a small disturbance at the level $10^{-17}$ to the initial conditions in Table~\ref{table:initial}, we  gain a non-periodic orbit that departs considerably from the original non-periodic ones for a long enough time.  Thus, the orbits given by the seven initial conditions  in Table~\ref{table:initial} are {\em unstable}.

Using the original initial conditions (in 5-digit precision) of the considered seven orbits reported by \v{S}uvakov and Dmitra\v{s}inovi\'{c} \cite{3body-2012}, we gain the same conclusion: the seven corresponding orbits are non-periodic and unstable.


 Therefore, according to our reliable numerical simulations based on the CNS,  either the initial conditions in Table~\ref{table:initial} of the reported seven ``periodic'' orbits are {\em not} accurate enough to predict a periodic ones, or the corresponding orbits are unstable.  Thus,  at least {\em seven} ``periodic''  orbits (listed in Table~\ref{table:initial}) of the Newtonian three-body problem found currently by \v{S}uvakov and Dmitra\v{s}inovi\'{c} \cite{3body-2012} are doubtful and should be checked very carefully.


The periodicity and stability  of the other orbits  reported  by \v{S}uvakov  and Dmitra\v{s}inovi\'{c} \cite{3body-2012} also should be doubt checked in details.

\section*{Ackowledgements}
The authors would like to express their sincere acknowledgements to Dr.  M. \v{S}uvakov  for providing us the more accurate initial conditions (listed in Table~\ref{table:initial}) of the 15 periodic orbits found in \cite{3body-2012}.  Thanks to Dr. Viswanath for providing us the more accurate initial conditions (in 500 significant digit) for the periodic solutions of Lorenz equation reported in \cite{Viswanath2004}.  This work is partly supported by National Natural Science Foundation of China under Grant No. 11272209.

\bibliography{3body}

\end{document}